\documentclass[amsart,amsmath,amssymb,onecolumn]{revtex4-2}

\pdfoutput=1
\usepackage{lipsum}
\usepackage[utf8]{inputenc}
\usepackage{graphicx}
\usepackage{float,placeins}
\usepackage[caption=false]{subfig}
\usepackage{chemarr,textcomp}
\usepackage{times,longtable}
\usepackage{hyperref}
\usepackage{color,multirow}
\usepackage{afterpage,enumitem}
\usepackage{soul,comment}
\usepackage[usenames,dvipsnames]{xcolor}
\usepackage{colortbl}
\usepackage{pbox}
\usepackage{capt-of}

\usepackage{subfig}

\usepackage{flushend}
\graphicspath{{large/}}
\bibpunct{[}{]}{,}{n}{}{,}

\definecolor{n1}{rgb}{1,0.7,0}
\definecolor{n23}{rgb}{0.8,0,0}
\definecolor{n45}{rgb}{0,0,0.8}
\definecolor{n6}{rgb}{.2,0.75,0.2}
\definecolor{n7}{rgb}{0.3,0.8,0.3}

\definecolor{m1}{rgb}{1,0,0}
\definecolor{m2}{rgb}{.8,0,.8}
\definecolor{m3}{rgb}{.3,0,1}
\definecolor{m4}{rgb}{.8,0.5,0}
\definecolor{m5}{rgb}{.5,.8,0}

\newcommand{\avg}[1]{\left\langle #1 \right\rangle}

\usepackage{cancel}
\usepackage{chemarr}
\usepackage{epsf,mathtools}
\usepackage{graphicx}
\usepackage{dcolumn}
\usepackage{bm}
\usepackage{amsmath,amssymb}
\usepackage{algorithm}
\usepackage{algpseudocode}
\usepackage[normalem]{ulem}

\begin{document}
\title{Evolutionary emergent metabolic interactions in cell cultures: A Statistical Mechanics point of view \footnote{Published in \textit{Physical Review E}, Vol. 112, Article 014402 (2025). DOI: 10.1103/dm8r-hzmk}}

\author{A.R. Batista-Tom\'as \thanks} \thanks{Corresponding author}
\email{albertbatista1988@gmail.com}
\affiliation{Group of Complex Systems and Statistical Physics, Department of Applied Physics, Physics Faculty. University of Havana, Cuba}

\author{C. D\'{\i}az-Faloh}
\affiliation{Group of Complex Systems and Statistical Physics, Physics Faculty. University of Havana, Cuba}

\author{R. Mulet}
\affiliation{Group of Complex Systems and Statistical Physics, Department of Theoretical Physics, Physics Faculty. University of Havana,  Cuba}

\begin{abstract}
Cell cultures exhibit rich and complex behaviors driven by dynamic metabolic interactions among cells. In this work, we present a model that captures these interactions through a framework inspired by statistical mechanics. Using Monte Carlo simulations, we explore the equilibrium and dynamical properties of a population of cells arranged in a two-dimensional lattice, where each cell is characterized by fluxes of three reactions: glucose consumption ($g$), respiration ($r$), and waste production/absorption ($w$). The system minimizes an energy function influenced by competitive ($J_g > 0$) and cooperative ($J_w < 0$) couplings between cells. Our results reveal three distinct phases: a competitive phase dominated by glucose competition, a cooperative phase marked by ordered waste exchange, and a disordered phase with local-scale cooperation. By incorporating evolutionary dynamics, we demonstrate how initially non-interacting cells can develop effective metabolic interactions, leading to heterogeneous cultures sustained by cross-feeding. These findings are further supported by analytical solutions derived using mean-field approximations. The model provides insights into how environmental constraints and stochastic fluctuations shape community structures, offering a versatile approach to study several emergent phenomena in biological systems.
\end{abstract}

\maketitle

\section{Introduction}

 Metabolic heterogeneity—the variability in metabolic states across individual cells—has emerged as a critical area of study with profound implications across diverse scientific domains. In oncology, it offers insight into therapeutic vulnerabilities and clinical prognoses\citep{Chang2015,Robertson-Tessi2015,Badr2020,Shirshin2022}. In biotechnology, it presents challenges for industrial efficiency\citep{Fernandez-de-Cossio2017,Pérez-Fernández2021}; and in microbial ecology, it supports competitive and cooperative dynamics\citep{Batista-Tomás2021}. Although these fields have traditionally approached heterogeneity from different perspectives, a unifying theme has emerged: metabolic diversity arises not only from cell-autonomous processes, but also from dynamic interactions between cells and their environment. Recent advances in single-cell technologies have further highlighted how these interactions evolve over time \citep{Onesto2023}, challenging traditional modeling paradigms that assume static metabolic networks.


The origins of metabolic heterogeneity are multifaceted. \textit{Intrinsic factors} such as cell-cycle fluctuations, aging, and transcriptional stochasticity generate baseline variability \citep{Kiviet2014,Takhaveev2018}. In contrast, \textit{extrinsic factors} such as environmental stress, nutrient shifts, and cell-cell interactions amplify this diversity \citep{Wimpenny2000,DiGregorio2016,Simsek2018,DeFrancesco2018,Kedia-Mehta2019,Ollé-Vila2019}. Among these, intercellular communication has emerged as a pivotal modulator. For example, yeast populations develop cooperative metabolic networks via metabolite exchange under stress \citep{Traven2012,DeMartino2016}, while tumors exploit glucose competition to suppress T-cell activity \citep{Chang2015}. Such interactions can induce persistent metabolic reprogramming, as seen in cancers where nutrient deprivation triggers irreversible pathway activation \citep{Wimpenny2000,DiGregorio2016}. Strikingly, tumors exhibit both Warburg and reverse Warburg effects simultaneously through metabolic coupling, allowing cross-feeding that fuels proliferation \citep{Wilde2017,Li2019,Liang2022}. Recent single-cell flux analyzes further quantify these dynamics, revealing structured proton-exchange networks \citep{DeMartino2018}. However, existing theoretical frameworks struggle to reconcile two key observations: (1) metabolic interactions are often plastic, rewiring rapidly in response to local conditions, and (2) these dynamic changes can lead to persistent population-level heterogeneity through feedback loops that alter both cellular states and their interaction rules.

Current modeling approaches fall into two broad categories. Constraint-based population models (CBPMs) extend traditional metabolic network analysis to cell communities but typically fix interaction rules over time\citep{Varma1993}. On the othr hand, ecological models account for environmental feedback, but often treat interaction modalities as static parameters rather than dynamic variables. A recent work has shown how intracellular metabolic states can alter environmental conditions which in turn affect cellular behavior \citep{Tahmineh2020}, but this models still assume that the fundamental rules governing cell-cell interactions remain unchanged. This limitation becomes particularly apparent in systems like tumors, where metabolic coupling between cancer cells and their microenvironment can shift abruptly between competitive and cooperative regimes, or in microbial communities where cross-feeding networks reorganize on timescales comparable to metabolic adjustments.  Pioneering work incorporated metabolic interactions using methods from disordered systems theory \citep{Dotsenko1995,Fernandez-De-Cossio2019,Fernandez-De-Cossio2020}, while later studies formalized how diffusion-limited exchanges constrain viable metabolic states \citep{DeMartino2017,Narayanankutty2024}. Yet a critical gap remains: existing models largely assume static interaction networks, neglecting how environmental or cellular feedback reshapes metabolic coupling over time.


Here, we bridge this gap by developing a theoretical framework that explicitly incorporates dynamic interaction rules into constraint-based modeling of cell populations keeping a description consistent with the usual framework of constraint based models. Our approach differs from previous work in three crucial aspects. First, we treat metabolic interactions as state-dependent functions that evolve on the basis of local metabolite concentrations and cellular metabolic states. Second, we integrate techniques from disordered systems theory with evolutionary dynamics to predict how transient environmental changes can drive populations toward stable heterogeneous states. Finally, our model reveals how interaction plasticity, the ability of cells to modify their coupling rules, can become an evolvable trait that shapes the resilience of the population. 


The manuscript is structured as follows: Section \ref{sec:model} details our modeling approach. Section \ref{sec:sim} presents numerical experiments replicating empirical conditions. Section \ref{sec:theo} adapts analytical techniques from disordered systems to evolving cell populations. Section \ref{sec:disc} evaluates implications, limitations, and future directions of our work. We conclude with broader insights into metabolic heterogeneity as an evolvable trait.

\section{Model}  
\label{sec:model}

Our model is built on four key assumptions, three of which were introduced in \citep{Fernandez-De-Cossio2019}: i) All cells share the same metabolic network, fully defined by stoichiometric constraints. ii) Nearby cells couple their metabolism, and the system minimizes a cost function (or energy) that accounts for these interactions. iii) Cells exhibit diverse metabolic states, described by a probability distribution. The fourth assumption, first introduced in this work, establishes a new dynamics for the couplings. That is, iv) the strength of interactions between neighboring cells is modulated by environmental signals. A detailed discussion of the advantages and limitations of these assumptions is provided in Section \ref{sec: dyn_model} and Section \ref{sec:disc}.

\subsection{Metabolic Network and Stoichiometric Constraints}
To simplify, we assume each cell operates a metabolic network that consumes a single substrate $S$ from the environment (glucose, for example) at a rate $g$. The substrate is transformed into an intermediary $P$ (pyruvate) obtaining energy $E$ (e.g, ATP), corresponding to the glycolysis process. After that, the intermediary can be either transformed into $E$ through respiration at a rate $r$, or secreted as waste by-products ($W$) with a rate $w$. These by-products can be lactate, acetate, or any alcohol derived from fermentation. Cells can also utilize $W$ as an alternative substrate, producing energy through respiration ($r$) via the lactate shuttle. The simplified metabolic network is shown in Figure \ref{schematic} (a). 

In the following, the sign of a particular flux will be negative if the reaction in the network is incoming at a node, and positive if it is outgoing. Therefore, in the model, the reaction flux $g$ is negative, while $r$ is positive. Note that $E$ is always produced, while $S$ is consumed. Also, thermodynamic reversibility imposes restrictions on the reaction fluxes inside the cell. Then: 
\begin{equation}\label{constr}
	\textbf{lb}_g\le g\le 0\;\;\;,\;\;\;0\le r\le \textbf{ub}_r
\end{equation}
where $\textbf{lb}_g$ and $\textbf{ub}_r$ are the lower and upper bound for reactions $g$ and $r$ respectively. In this work, we set $\textbf{lb}_g=-1$ and $\textbf{ub}_r=1$, in order to simplify the calculations. 

Balance equation for the intermediate $P$ leads to  the stoichiometric constraint:
\begin{equation}\label{stech}
	-g = r + w,
\end{equation}
where $-1 \le g \le 0$, $0 \le r \le 1$. These constraints imply $-1 < w < 1$, which properly describes the possibility of waste absorption or secretion.

\subsection{Energy Minimization and Interacting Cells}
In the condition of cell independence, it can be defined an utility function, related to the biomass production (or ATP in its case), which can be expressed in some cases as a linear combination of the reaction fluxes \citep{Fernandez-De-Cossio2020}. In this simple scenario, each cell $i$ maximizes the utility function:
\begin{equation}
	-E_i = h_g g_i + h_r r_i,
\end{equation}
where $h_g$ and $h_r$ represent the contributions of glycolysis- and respiration to biomass production, respectively ($h_w=0$). The minus sign makes the equivalence of minimizing the energy (or cost) function $E_i$ of each cell, which gives a direct interpretation in the framework of statistical mechanics.

While traditional Constraint-Based Modeling (CBM) assumes cells independently maximize their utility functions, we introduce interactions between cells by minimizing a global energy function $H$:
\begin{equation}
	H = \sum_i E_i + V,
\end{equation}
where $V$ represents the energy associated to metabolic interactions between neighboring cells, contributing to the biomass production in a similar way as the utility function. Then $V$ can be seen as the contribution to the utility function due to interactions between neighbors, weighted through their couplings. Specifically, we consider two types of interactions:
\begin{equation}
	V_g = - \sum_{i,j} J_{gij} g_i g_j \quad \text{and} \quad V_w = - \sum_{i,j} J_{wij}w_i w_j.
\end{equation}
Here, the couplings $J_{gij} > 0$ model competition for glucose between neighbors $i,j$, favoring states where $g_i$ and $g_j$ have the same sign. Conversely, $J_{wij} < 0$ favors cooperation, where one cell's waste (e.g., lactate) becomes another's energy source. In the general case, $J_{gij}$ and $J_{wij}$ are different for each pair of neighbors. This framework generalizes the Linear Programming hypothesis, often applied to bacterial colonies.

\subsection{Metabolic States and Boltzmann Distribution}
We can define the equilibrium metabolic state of the culture with the set $\underline{g}, \underline{r}, \underline{w}$, where $\underline{g}=\{g_1,g_2,...,g_N\}$, and the same for the other reactions. In equilibrium, the metabolic states distribution of cells in a culture follows a Boltzmann distribution:
\begin{equation}
	p(\underline{g}, \underline{r}, \underline{w}) \propto e^{\beta \sum_{s} \left( \sum_{i<j} J_{sij} s_i s_j + \sum_i h_i s_i \right)} \prod_{i=1}^N \delta(g_i - r_i - w_i),
	\label{eq:Boltzmann}
\end{equation}
where $s$ denotes the different reactions, $\beta$ quantifies heterogeneity of the culture, and the stoichiometric constraint is enforced by $\delta(\cdot)$. In the limit $\beta \to \infty$, the system converges to a single minimum, corresponding to the linear programming solution of the Flux Balance Analysis. For $\beta \to 0$, $p(\underline{g}, \underline{r}, \underline{w})$ becomes uniformly distributed in all configurations satisfying $-g = r + w$. In this form, equation (\ref{eq:Boltzmann}) can be interpreted as the canonical distribution of $N$ continuous coupled spins, subject to an external field $h$ with inverse temperature $\beta$ \citep{Fernandez-De-Cossio2019,Fernandez-De-Cossio2020}.

\subsection{Couplings Dynamics}\label{sec: dyn_model}
Experimental evidence suggests that the strengths of the interaction between cells adapt to environmental changes. For example, in \textit{Saccharomyces cerevisiae} cultures, cells on the periphery exhibit a higher glycolysis activity, while central cells favor respiration due to glucose depletion \citep{Traven2012}. More generally, two mechanisms for this adaptation are possible, first, daughter cells that are more resistant (or fit) to new environmental conditions have a larger possibility to survive, second, regulatory mechanisms can up- and down-regulate different enzymes, changing the possible metabolic phenotype of the cell. Although our model does not distinguish explicitly between these mechanisms, it reflects the richness of the possible dynamical behavior of these systems. A comprehensive review of cell communication mechanisms is provided in \citep{Armigol2020}.

To model this, we adapt a framework from neuronal learning \citep{Coolen1993,Penney1993,Dotsenko1994,Dotsenko1995}. The couplings $J_{gij}$ and $J_{wij}$ evolve according to:
\begin{equation}
	\tau \frac{d}{dt} J_{sij} = \frac{1}{N} \left\langle s_i s_j \right\rangle_{\{J_{sij}\}} - \mu (J_{sij} - J_s/N) + \frac{1}{\sqrt{N}} \eta_{sij}(t),
	\label{eq:evolJ}
\end{equation}
where $s \in \{g, r, w\}$. The Gaussian noise term $\eta_{sij}(t)$ has zero mean and correlations:
\begin{equation}
	\left\langle \eta_{sij}(t) \eta_{sij}(t') \right\rangle = \frac{2\mathcal{T}}{\tilde{\beta}} \delta_{{sij},{skl}} \delta(t - t').
\end{equation}

The first two terms of the right-hand side of Eq. (\ref{eq:evolJ}) correspond to the deterministic drift of the couplings. They capture the evolution of couplings based on neighbors' correlations (first term), and the evolution towards a ground value $J_s$ (second term). This value may vary between different types of cells. For example, neurons have a different intrinsic coupling compared to other tissues. Also in this term, $\mu$ establishes the rate at which the system evolves towards $J_s$. It can be used to understand the relation between the tissue's natural drift and the drift due to the correlation between neighbors.

The timescale $\tau$ is the time covariance of the noise, and its value determines the dominance of the deterministic or the stochastic part of Eq.(\ref{eq:evolJ}) \citep{Haunggi1994}. On the other hand, the inverse temperature $\tilde{\beta}$ is related to the stochastic dynamics of the couplings, and its relationship to the cellular system's $\beta$ determines the system's equilibrium state. When the noise is small, i.e. $\tilde{\beta} \to \infty$, the deterministic terms dominate the dynamics. On the other hand, if $\tilde{\beta} \to 0$, the dynamics becomes fully stochastic and the noise dominates the dynamics. This separation of timescales—where coupling dynamics operate slower than internal fluxes—aligns with biological observations \citep{Coolen1993,Traven2012,DeMartino2016}. In our case, we set $\tau=1$ for simplicity, while controlling the time-scale separation between the two stochastic processes with the ratio $n=\tilde\beta/\beta$.

Summarizing, the rule that governs the evolution of the couplings $J_{sij}$, as well as the proper selection of the model parameters, can be understood as an effective modeling strategy that captures a variety of underlying processes—both regulatory and evolutionary—which shape the system's behavior. This allows the model to reproduce well-known biological phenomena such as the Warburg effect, inverse Warburg effect, or synaptic learning driven by impulse activity \citep{Jonker1991,Jonker1993}.

The dynamical equations describing the evolution of the interactions lack of experimental support, but represent a plausible biological scenario and serve as a first proxy to justify the existence of such evolutionary behavior in the culture. We chose them to exploit our previous understanding of this kind of dynamics, while other evolutionary processes can be adapted well to our formalism. Therefore, the framework can also be adapted to describe other scenarios, such as noise-free evolutionary models \citep{Shinomoto1987}, which correspond to the limit $\beta\to \infty$, or cases governed by anti-Hebbian learning rules \citep{Jonker1993}, which would involve changing the sign of $\avg{s_is_j}_{J_{sij}}$. In this way, Equation \ref{eq:evolJ} should be viewed as a simplified yet flexible tool for modeling potentially complex and multicausal real-world dynamics.

\section{Simulations}
\label{sec:sim}

This section presents the results of Monte Carlo (MC) simulations for the model described above. Our objectives are twofold: first, to illustrate the rich phenomenology that emerges from the model's underlying hypotheses; and second, to highlight its versatility, which makes it suitable, at least in principle, for describing real-world experimental scenarios.

\begin{figure}[b!]
	\centering
	\includegraphics[width=0.80\textwidth]{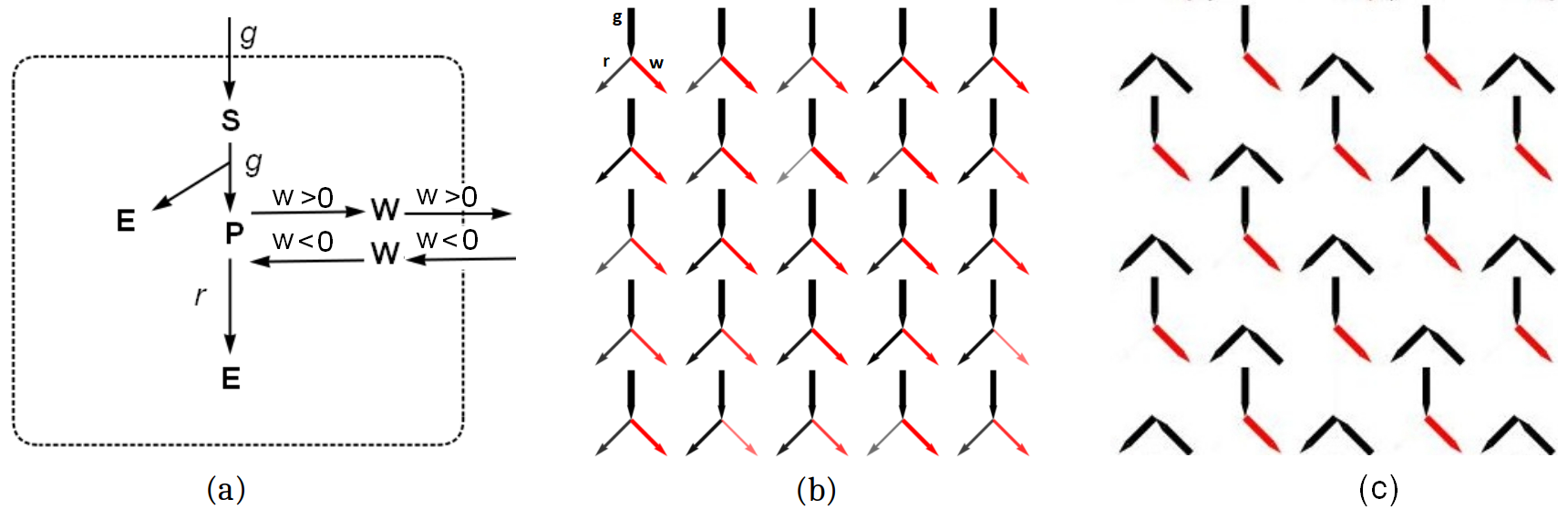}
	\caption{(a) Representation of the simplified metabolic network. Arrows represent the metabolic fluxes, and the capital letters represent the involved metabolites. (b) Schematic representation of the bidimensional lattice for a competitive scenario. (c) Schematic representation of the bidimensional lattice for a cooperative scenario. Collaboration is evident between the cells. The top arrow corresponds to flux $g$, while the bottom left arrow and the bottom right arrow correspond to $r$ and $w$, respectively. In all cases, arrows pointing to the center are representative of a negative flux, and arrows pointing out represent a positive flux. }
	\label{schematic}
\end{figure}

\subsection{Case of Equilibrium}

We begin by reviewing, and extending, the equilibrium properties of the model without the couplings dynamics, to establish a connection between the original findings in \citep{Fernandez-De-Cossio2019, Fernandez-De-Cossio2020} and the dynamical model introduced here. Unlike previous works, where only couplings in the waste reaction ($w$) were considered, we assume that couplings also affect the reaction $g$ (competition). This extension allows us to explore how the culture's possible states depend on the coupling strengths.

In the following, all simulations are performed on a two-dimensional lattice, with interactions limited to nearest neighbors. While we acknowledge that this choice may not be optimal in certain cases—such as cells in suspensions it suits very well to describe biofilms, or simple tissues.

We can define $m_s$ as the average rate per cell of reaction $s$ across the system. At very low values of $\beta$, the exponential term in the Boltzmann factor becomes negligible, and the equilibrium state is dictated solely by the constraints. In this case, $-m_g = m_r = 0.5$ and $m_w = 0$. At high values of $\beta$, the exponential term dominates. In the absence of interactions (i.e., $J_{gij} = J_{wij} = 0$ for all $(i,j)$), each cell maximizes its glucose uptake and fully respires it, resulting in $-m_g = m_r = 1$ and $m_w = 0$.

When interactions are present, the behavior of the system becomes more complex. We consider the case where $J_{sij}$ values are drawn from a Gaussian distribution with means $J_w < 0$ and $J_g > 0$, and variances $\Delta = 1$. This generalizes the model in \citep{Fernandez-De-Cossio2020}, where only couplings via the waste product $w$ were considered.

For any finite $\beta$, if $|J_{g}|$ is sufficiently larger than $|J_w|$, we expect a nonzero average secretion (or absorption) of the waste product $w$, i.e., $|m_w| > 0$. In this scenario, maximizing $g$ for each pair of cells takes precedence over other reactions. As a result, cells compete with their neighbors for glucose, producing $r$ and $w$ according to the interaction terms.

This competitive behavior is confirmed in our MC simulations, as shown in Figure \ref{schematic}(b). In this scenario, the interpretation is that all cells are on average in the same metabolic state (competitive in this case). In the visual representation of fluxes in Figure \ref{schematic} (b)(c) arrows represent reaction fluxes, with absorption (negative flux) depicted as an incoming arrow and secretion (positive flux) as an outgoing arrow. The thickness of the arrows corresponds to the magnitude of the fluxes. The upper arrow represents glucose consumption ($g$), while the left and right arrows correspond to respiration ($r$) and waste production/absorption ($w$), respectively.

\begin{figure}[!h]
	\centering
	\includegraphics[width=0.95\textwidth]{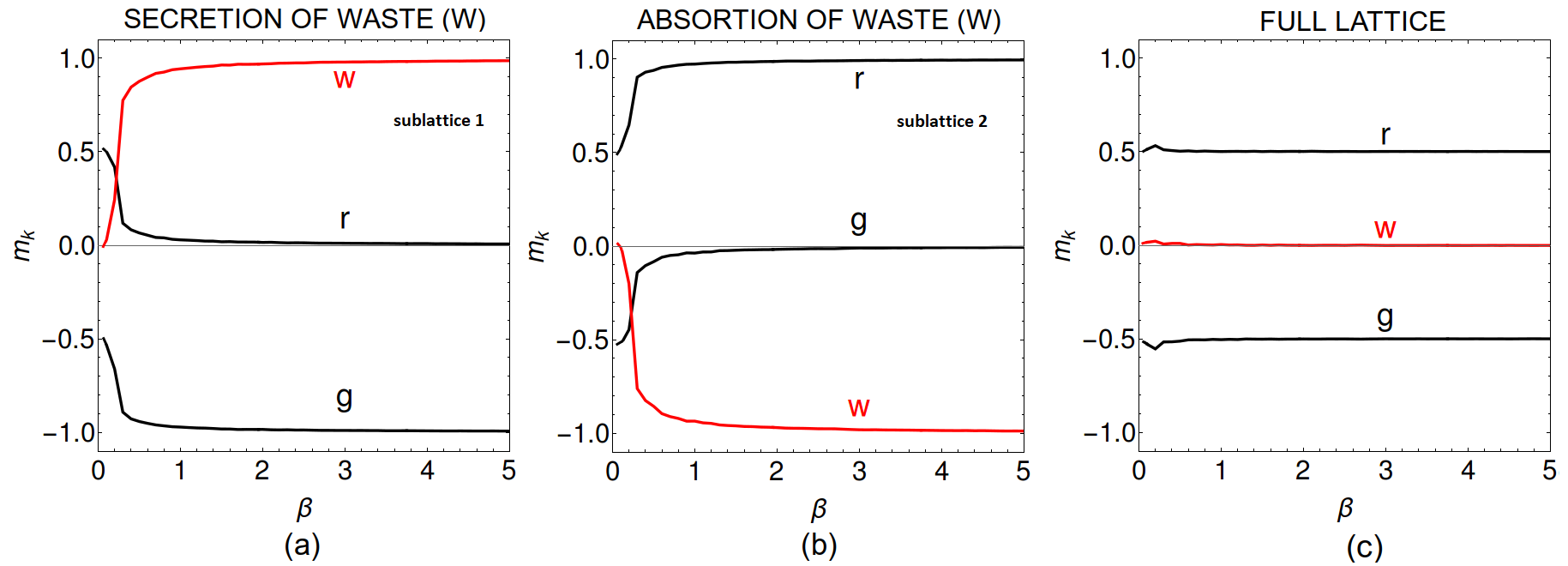}
	\caption{(a) Staggered magnetization for each value of $\beta$ when $J_w = -10$, $J_g = 2$, and $\Delta = 1$. The panel represents cells that take glucose from the environment and secrete it as a by-product, while (b) represents cells that use $W$ as a substrate for respiration. (c) Magnetization across all cells.}
	\label{staggered}
\end{figure}

Conversely, when $|J_w|$ increases, the exchange of secreted by-products becomes significant, altering the culture's behavior. In this case, a collaborative state emerges, as shown in Figure \ref{schematic}(c). In the bidimensional lattice, neighboring cells maximize their fluxes $w$, with one cell absorbing and the other secreting. It means that at this point, collaboration via lactate shuttle is playing an important role in the culture's behavior. This collaboration creates two distinct sub-lattices at the microscopic level, as illustrated in Figure \ref{staggered}(a)(b). Some cells (sublattice 1) take glucose from the environment and secrete it as a by-product $W$ ($w>0$), while others (sublattice 2) use $W$ as a substrate for respiration ($w<0$). Full-system averages obscure this microscopic behavior, as shown in panel (c). As $\beta$ increases, the ordered configuration becomes dominant, while lower values of $\beta$ introduce randomness, disrupting the collaboration.

\begin{figure}[h!]
	\centering
	\includegraphics[width=0.4\textwidth]{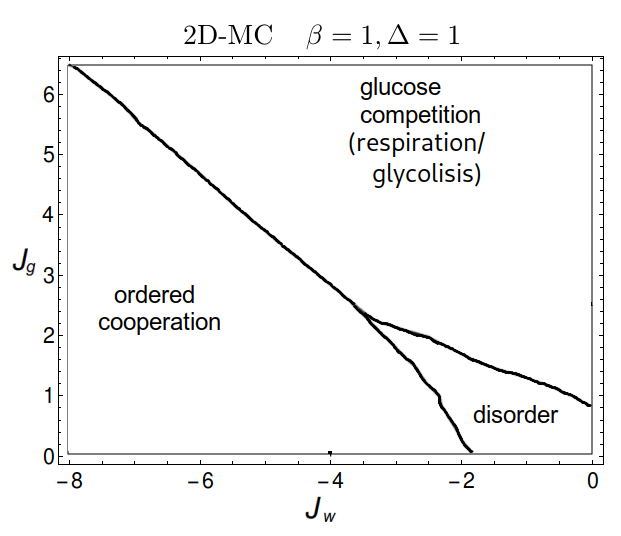}
	\caption{Phase diagram obtained from the bidimensional model with respect to the couplings $J_g$ and $J_w$, for $\beta = 1$.}
	\label{fases}
\end{figure}

To generalize, Figure \ref{fases} shows a $J_g$-$J_w$ phase diagram of the system for $\beta = 1$, obtained from extensive MC simulations. Depending on the values of $J_g$ and $J_w$, three distinct phases can be identified:
\begin{itemize}
    \item The \textit{competitive phase}, marked by positive magnetization in $w$, resulting from competition for glucose among neighboring cells.
	\item The \textit{cooperative phase}, characterized by a well-ordered structure similar to Figure \ref{schematic}(c).
	\item The \textit{disordered phase}, arising from conflicting interactions and marked by the formation of small clusters of cooperating cells, indicates local-scale cooperation. These emerging collaborative phenomena are indicative of the possible emergence of the reverse Warburg effect in tissue.
\end{itemize}

\subsection{Dynamical Simulations}
\label{ssec:dyn}

In the previous section, we described the equilibrium phenomenology of the model. Here, we demonstrate the relevance of this phenomenology by showing that simple evolutionary dynamics can drive a system of initially non-interacting cells (i.e., all $J_{sij} = 0$) toward a state of effective metabolic interactions. The underlying idea is that cells capable of exploiting their environment behave better than those that do not. For instance, cells that consume glucose more efficiently outperform others and dominate the culture. Alternatively, when cooperating, cells that utilize metabolic by-products from neighboring cells as resources can survive even with reduced glucose consumption. These mechanisms introduce evolutionary forces that may favor the emergence of metabolic interactions.

To incorporate these dynamics into our Monte Carlo (MC) framework, we consider two processes running in parallel: one governing the change of fluxes within each cell and another governing the evolution of the interactions. Each process operates on its own timescale and temperature. 

The change of fluxes follows the Metropolis rule, defined by:
\begin{equation}
	\Delta H(s) = \sum_{s,i} \left( h_s s_i - h_s s'_{i} \right) - \sum_{s,i<j} \left( J_{sij} s_i s_j - J_{sij} s'_{ri} s_j \right),
	\label{eq:delta-nu}
\end{equation}
where $s'$ represents a proposed update to the flux configuration. The acceptance probability depends on the change in energy $\Delta H(s)$.

\begin{equation}
	\Delta \mathcal{H} = \frac{1}{2} \mu \cdot 4 \left( {J'_{ij}}^2 - {J_{ij}}^2 \right) - \frac{1}{\beta} \ln \frac{Z(J'_{ij}, [J_{/ij}])}{Z(J_{ij}, [J_{/ij}])},
	\label{eq:delta_J}
\end{equation}
where the factor $4$ accounts for nearest-neighbor interactions in the simulations. Here, $Z(J_{ij}, [J_{/ij}])$ is the partition function of the current state, with no changes to the $J_{ij}$ values, while $Z(J'_{ij}, [J_{/ij}])$ is the partition function if the proposed change from $J_{ij}$ to $J'_{ij}$ is accepted.

\begin{figure}[t!]
	\centering
	\includegraphics[width=0.9\textwidth]{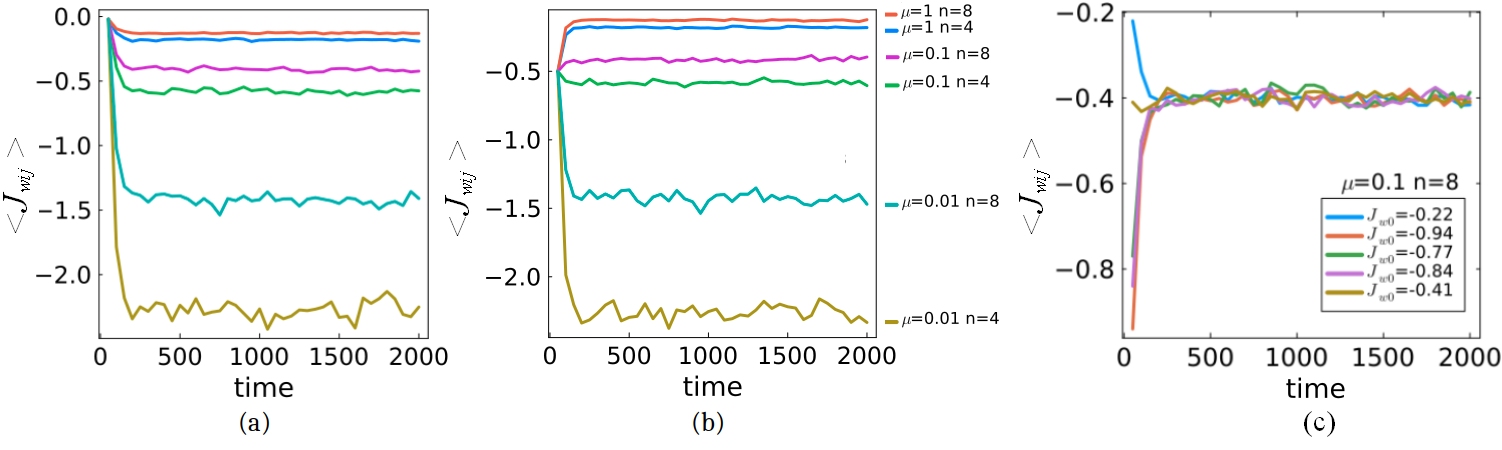}
	\caption{The curves represent the MC averages of the couplings over the population: $\avg{J_{wij}} $. Simulations were performed for different values of $n$ and $\mu$, with two different initial conditions. Fields $J_s$ are zero in all cases. Initial values are: (a)  $J^0_{wij} = -0.02$ and (b) $J^0_{wij} = -0.50$ . In both graphs $\mu = 1, 0.1, 0.01$; $n = 4, 8$. The curves, from top to bottom, correspond to the legend from top to bottom as well. (c) Dependence on the initial conditions $(J^0_{wij})$. In all cases the results were averaged over 10 simulations.}
	\label{Jeq}  
\end{figure}

Figure \ref{Jeq} shows the results for the coupling evolution $\avg{J_{wij}}$ at different values of $\mu$ and $n$. The system is initialized with different values of $J_{wij}$, while $J_g$ is kept at zero throughout the simulation, as we focus on the cooperative scenario. When the system reaches equilibrium, larger values of $J_{wij}$ increase the likelihood of stabilizing in a cooperative state, in consistence with the phase diagram shown in Figure \ref{fases}. Additionally, lower values of $\mu$ or $n$ drive the culture toward a more cooperative outcome. This supports the notion that a cooperative stationary state emerges when interaction dynamics evolve more slowly than flux dynamics but are still fast enough to propagate within the culture. Furthermore, Figure \ref{Jeq}(c) demonstrates that the initial conditions of the interactions have minimal influence on the equilibrium state. 

\subsection{Isolated Cells}

To demonstrate the versatility of our approach, we simulate an experiment reported in \citep{Traven2012}, where a yeast culture is placed in an environment where only the outermost cells have access to substrate. Under these conditions, it was observed that the outermost cells consumed large amounts of substrate and secreted a byproduct, while the inner cells began utilizing this byproduct to produce ATP. This is an excellent example of how, under specific environmental conditions, a culture can evolve to adopt the most advantageous metabolic configuration for the system as a whole. This scenario aligns naturally with our model.

\begin{figure}[b!]
	\centering
	\includegraphics[width=0.90\textwidth]{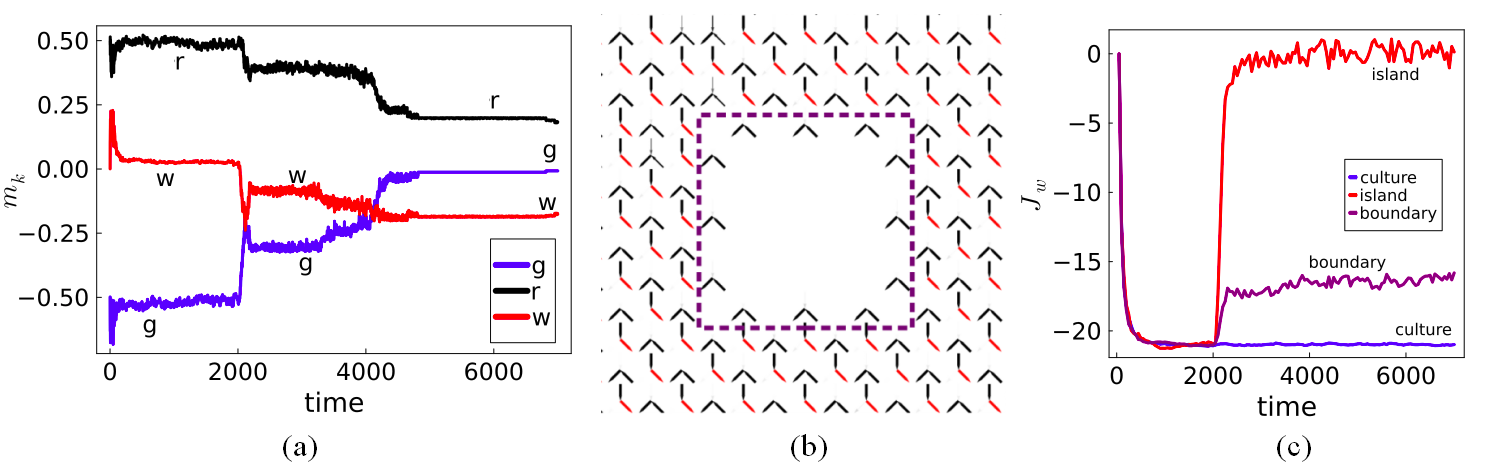}
	\caption{(a) Average fluxes $g$, $r$, and $w$ in the $7 \times 7$ isolated subset. Glucose begins to deplete for some cells at $t = 2000$ and is completely exhausted around $t = 4000$. Cells are forced to utilize $w$ to survive. (b) 2D lattice representation of the cell culture, with a $7 \times 7$ subset in the center subjected to finite glucose conditions. In this snapshot, glucose has already been depleted in this subset, and the only surviving cells are those at the boundary, which can take up $w$ from the surrounding culture.  
		(c) Average $J_w$ values for the entire culture, the subset, and its boundary. Once glucose is depleted, the boundary cells maintain stronger interactions, while those in the center of the subset eventually ran out of resources. }
	\label{ais}
\end{figure}

In this case, we modify the flux limits to $-1 \leq g \leq 0$, $0 \leq r \leq 0.5$, and $-0.5 < w < 1$ to allow for the accumulation of the byproduct. Additionally, glucose is no longer supplied constantly within a small $7 \times 7$ subset of the culture, while the rest of the culture retains an unlimited glucose supply. In this subset, cells begin with a finite amount of glucose, which is consumed according to Michaelis-Menten kinetics. This initial glucose supply is sufficient to sustain the cells for a limited period. Once depleted, glucose becomes scarce, forcing the cells to adapt to the new environmental conditions. This adaptation is evident in Figure \ref{ais}(a), where the average fluxes $m_s$ are plotted for the $7 \times 7$ subset. Initially, these cells behave similarly to the rest of the culture, which has an unlimited glucose supply. However, once their glucose runs out, the island's inner cells' fluxes go to zero, which can be interpreted as if the cells died from nutrient deprivation \citep{DiGregorio2016}. Also, cells in the island's outermost layer can switch to consuming the byproduct ($w$) generated by the external cells, and use it to obtain energy by a lactate shuttle mechanism. This adaptation is reflected in the negative values of $m_w$ at the end of the simulation.

As in the experiment, the goal is to examine how cell interactions evolve in a changing environment. The simulations were performed while maintaining the interaction dynamics described earlier. As can be noted in Figure \ref{ais}(a) and Figure \ref{ais}(c). The dynamics for the couplings evolution of the cells in the island changes abruptly at $t=2000$, when the Michaelis-Menten process depletes the carbon reserves $g$ for some cells, (which has an initial stock of 2000, being consumed at a maximum rate of one unit per second). The process continues up to $t=4000$, where glucose is fully depleted for all cells in the island. This explains the sharp behavior of the curves at this particular times. The strong interactions developed within the culture reflect in the fact that the outmost layer of the $7 \times 7$ subset can consume the byproduct ($w$) secreted by the surrounding culture. However, since only nearest-neighbor interactions are considered, the flux of the cells that are not at the boundary, hence not in direct contact with the rest of the culture, eventually goes to zero. This, combined with the relatively low $w$ fluxes maintained at the boundary, causes $\avg{|J_{wij}|}$ to decrease. Nevertheless, it is still sufficient to sustain the fluxes on some cells, using lactate shuttle as its main metabolic pathway. These results are illustrated in Figure \ref{ais}, where panel (b)  shows the cell configuration on the 2D lattice, and panel (c) presents the evolution of the interactions within the culture, the isolated subset, and its boundaries.

\section{Analytical Solution}
\label{sec:theo}

In this section, we present an analytical solution of the model with dynamical evolution for couplings. To do that, the same approach as for Sherrington-Kirkpatrick spin glass systems was followed \citep{Sherrington1975}, although we considered continuous spins in the space of fluxes, instead of Ising spins \cite{Fernandez-De-Cossio2020}. Furthermore, introducing the new temperature for the coupling dynamics makes it necessary to use techniques of partially annealed systems, developed by \citep{Coolen1993, Penney1993, Dotsenko1995} . The complete derivation for the cell and coupling system is provided in the Appendix \ref{A1}.

We begin by noting that the Boltzmann ansatz for the probability distribution allows us to describe the cells as interacting through a Hamiltonian defined as:
\begin{equation}
	H(\{s\})_{\{J_{sij}\}} = -\sum_s \sum_i h_s s_i - \sum_s \sum_{i<j} J_{sij} s_i s_j,
	\label{hamiltonising}
\end{equation}
where the subscript indicates that the Hamiltonian is defined for a fixed set of couplings $\{J_{sij}\}$ at a given instant of the evolution. The thermodynamics of the cells system are then derived from the partition function:
\begin{equation}
	Z(\{J_{sij}\}) = \underset{\{s\}}{\text{Tr}} \left( e^{-\beta H(\{s\})} \right).
\end{equation}

To incorporate the coupling evolution described by Equation (\ref{eq:evolJ}), and defining $n = \tilde{\beta}/\beta$, the partition function for the full dynamics of the coupled system (cells and couplings) becomes:
\begin{equation}
	Z_{\tilde{\beta}} = \int dJ_{sij} \, Z^n(\{J_{sij}\}) \, e^{-\frac{1}{2} \tilde{\beta} \mu N \sum_{s,i<j} \left(J_{sij} - J_{s}/N\right)^2}.
	\label{Zfull}
\end{equation}
This partition function can be used to compute the moments of the dynamical variables in the usual manner \citep{Penney1993, Coolen1993}. Specifically, the first moment of the couplings to leading order $N$ after equilibrium is given by:
\begin{equation}
	\frac{N}{\tilde{\beta}} \frac{\partial \ln Z_{\tilde{\beta}}}{\partial J_{s}}  = \mu N \overline{J_{sij}} - \mu J_s = \sum_a (m_s^a)^2,
\end{equation}
where the index $a$ corresponds to the replica index (see Appendix \ref{A1} for the full derivation). 

For the moments of the spins, it is:
\begin{equation}
	\frac{1}{\tilde{\beta}} \frac{\partial \ln Z_{\tilde{\beta}}}{\partial h_i} \Big|_{h_i = h} = \overline{\langle s_i \rangle}.
\end{equation}
Here, $\langle \cdot \rangle$ represents the Boltzmann average with the Hamiltonian (\ref{hamiltonising}) for a fixed set of $J_{sij}$'s, while the overbar denotes the average over the asymptotic dynamics of the $J_{sij}$'s.

Using standard techniques from spin glass theory, we define the order parameters:
\begin{equation}
	m_s^a = \frac{1}{N} \sum_i s_i^a, \quad q_s^{ab} = \frac{1}{N} \sum_i s_i^a s_i^b,
\end{equation}
where $m_s^a$ is the first moment of the reaction flux $s$ over different replicas, and $q_s^{ab}$ is the overlap between the fluxes of a cell for different replicas. When $a = b$, $q_s^{aa}$ corresponds to the second moment of the reaction flux for one replica. In terms of a population, $m_s^a$ represents the average of the reaction flux $s$ over the cells, while $q_s^{ab}$ quantifies the overlap of $s$ for configurations at two widely separated times.

The parameter $n$ remains finite and integer (although a real analytic continuation can be performed), corresponding to a partially annealed system \citep{Coolen1993, Penney1993,Dotsenko1995}. After performing the typical algebraic manipulations, we apply the replica symmetric approximation, that is, $m_s^a = m_s$, $q_s^{ab} = q_s$ for $a \neq b$, and $q_s^{aa} = \zeta_s$. The free-energy density of the system is then:
\begin{equation}
	f(m_s, q_s, \zeta_s) = -\sum_s \left[ \frac{J_s m_s^2}{2} + \frac{\Delta_r^2}{4} \left( \zeta_s^2 + (n-1) q_s^2 \right) \right] + \frac{1}{n} \ln \int \left\{ \text{Tr } \left[ \exp(-\mathbb{H}_{RS}) \right] \right\}^n D\vec{t_s},
	\label{freeenergy}
\end{equation}
where:
\begin{equation}
	\mathbb{H}_{RS} = -\sum_s \left( \beta h_s + J_s m_s + \Delta_r \sqrt{q_s} t_s \right) s - \sum_s \frac{\Delta_r^2}{2} (\zeta_s^2 - q_s^2) s^2,
\end{equation}
and $\Delta^2 = \frac{1}{\tilde\beta \mu}$. The saddle-point solutions in the thermodynamic limit minimize the free energy. In the replica symmetric approximation, the interpretation of the order parameters is:
\begin{equation}
	m_s = \overline{\langle s \rangle}, \quad q_s = \overline{\langle s \rangle^2}, \quad \zeta_s = \overline{\langle s^2 \rangle}.
	\label{saddle}
\end{equation}
Here, $\langle s \rangle$ and $\langle s^2 \rangle$ are the first and second moments of the reaction fluxes, while $\langle s \rangle^2$ is the overlap between the reaction fluxes at two widely separated times.

Finally, it is important to note that the parameters in Equation (\ref{saddle}) are such that they minimize the free-energy function, and the Boltzmann average $\langle \cdot \rangle$ is performed with the $n$-replicated Hamiltonian $\mathbb{H}_{RS}$. The overbar $\overline{(\cdot)}$ denotes a Gaussian average over the variables $t_s$, defined as:
\begin{equation}\label{eq:Gaussavg}
	\overline{\langle \cdot \rangle} = \frac{\displaystyle \int D\vec {t_s} \; \left[ \text{Tr } \left( e^{-\mathbb{H}_{RS}} \right) \right]^n \left( \frac{\text{Tr } \left( e^{-\mathbb{H}_{RS}} (\cdot) \right)}{\text{Tr } \left( e^{-\mathbb{H}_{RS}} \right)} \right)}{\displaystyle \int D\vec{t_s} \; \left[ \text{Tr } \left( e^{-\mathbb{H}_{RS}} \right) \right]^n},
\end{equation}
where $D\vec {t_s} = \prod_s dt_s \, \exp(-t_s^2 / 2) / \sqrt{2\pi}$.

\subsection{Evolution Phase Diagram}
The analytical solutions obtained above bring the opportunity of exploring the full space of parameters and, at the same time, validate the dynamical model used for couplings evolution. Following the previous analysis, we separate the solutions for the two cases: flux equilibrium without and with evolution on the couplings.

\begin{figure}[htb]
	\centering
	\includegraphics[width=0.80\textwidth]{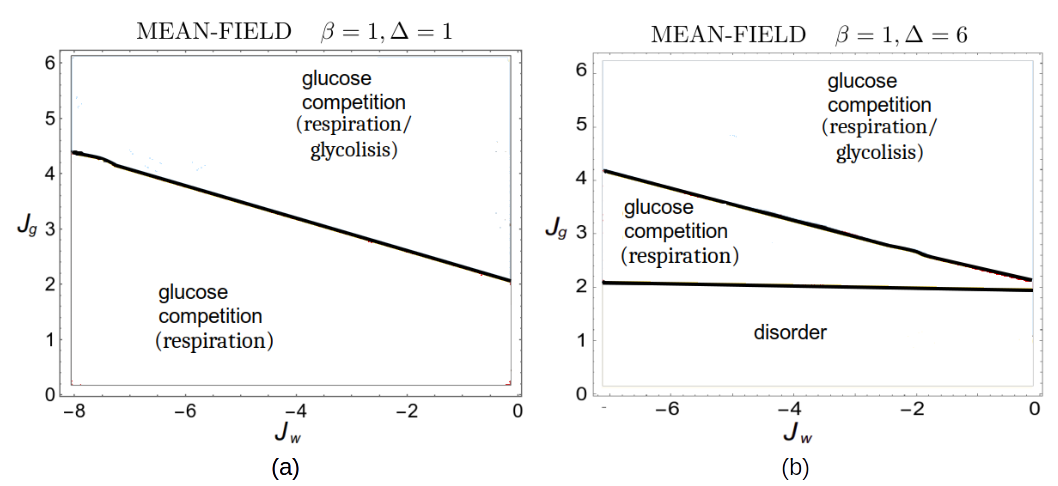}
	\caption{(a) Analytical mean-field solution for $\beta = 1, \Delta=1$. (b) Analytical solution for $\beta=1,\Delta = 6$.}
	\label{phase_MF}
\end{figure}

The first scenario, presented above in Figure \ref{fases} (for MC simulations), is now obtained using the theoretical mean-field solutions. The results, shown in Figure \ref{phase_MF}(a), reproduce the simulations for the competitive region. However, the ordered cooperation phase is not obtained, as a direct consequence of the limitations of the mean-field theory of spin glasses: it fails when most of the couplings $J_{wij}$ are negative. This is also confirmed in Figure \ref{phase_MF} (b), where a great part of the coupling $J_{wij}$ is positive (although its average is negative). In this case, the disorder phase region appears, for low values of $J_{g}$, replicating the disorder phase in Figure \ref{fases}. 

\begin{figure}[h!]
	\centering
	\includegraphics[width=0.9\textwidth]{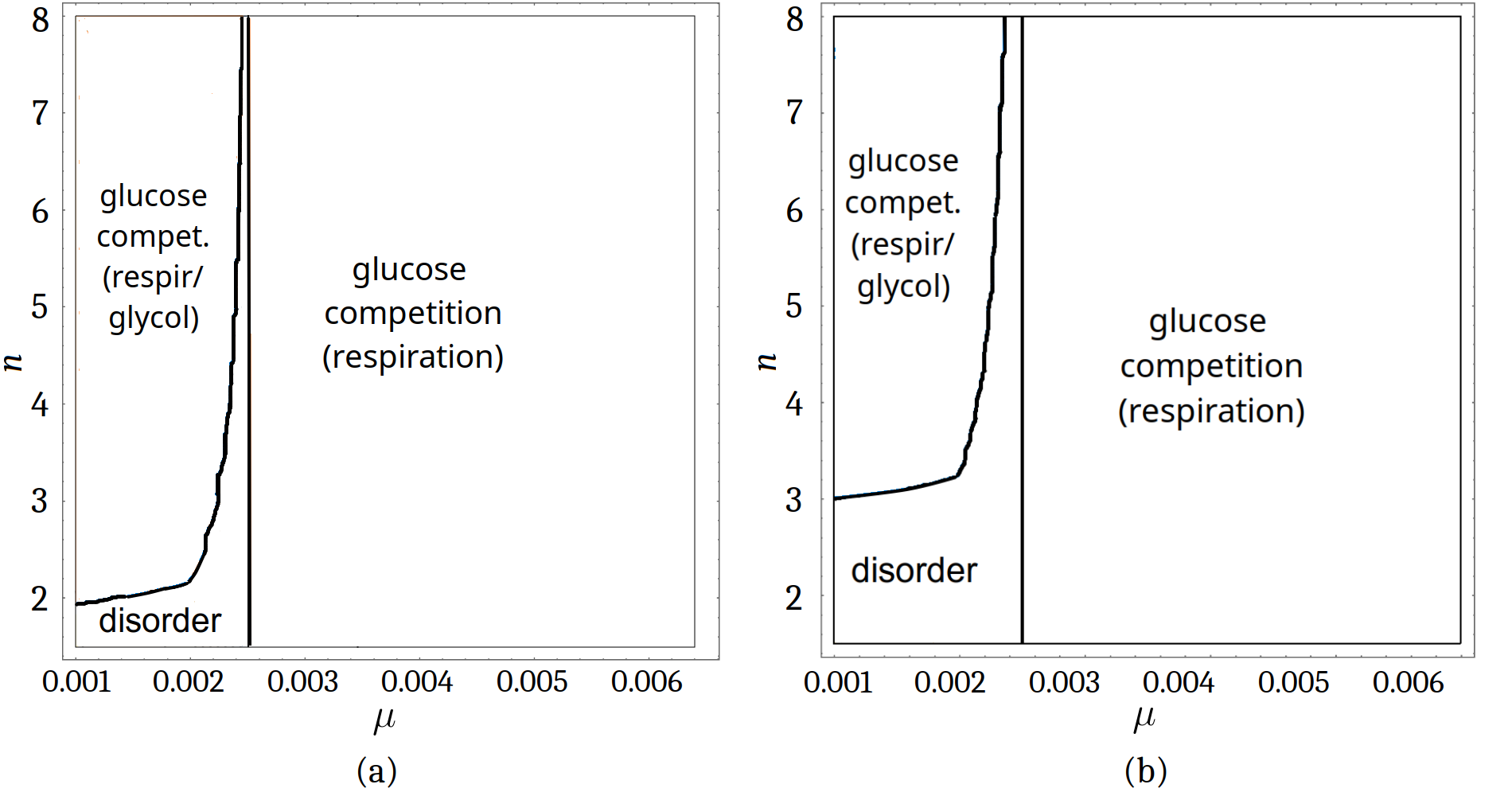}	
	\caption{(a) Phases exhibited by the culture for $J_w = J_g = 0$, $\beta = 0.05$, and the dynamic evolution (\ref{eq:evolJ}) applied to the third reaction. (b) For $J_w = -10$, a larger collaborative region (disordered phase) is observed in the phase space.}
	\label{last}
\end{figure}

Once obtained the results without evolution, we can follow the case with evolution as well. The results obtained in Section \ref{sec:sim} can now be generalized to analyze the phases adopted by the tissue as a function of the parameters $\mu$ and $n$. As mentioned in Section \ref{sec: dyn_model}, the parameter $\mu$ represents the rate of evolution to ground value $J_s$. On the other hand, $n$ determines whether the couplings evolution is governed by deterministic factors when $n$ is large, or the stochastic noise given by $\tilde{\beta}$ when $n$ is small. The resulting  equilibrium phases are shown in Figure \ref{last}.

Let us first discuss the phase diagram of Figure \ref{last}(a) in detail. At all times, the evolution of the couplings arises from the competition between three factors: the rate of evolution $\mu$ toward the fields $J_s$, the correlations between fluxes of neighboring cells modulated by $\beta$, and the stochastic noise determined by $\tilde{\beta}$. When the "natural drift" from tissue is very weak ($\mu \to 0$), the system's behavior depends on the relationship $n = \tilde{\beta}/\beta$. In this case:
\begin{itemize}
	\item For larger values of $n$, the variance $\Delta^2=1/\tilde\beta\mu$ of the coupling equilibrium distribution decreases, implying that all $J_{sij}$'s tend to be identical. This leads to a competitive configuration \citep{Sherrington1975, Penney1993, Coolen1993,Fernandez-De-Cossio2019, Fernandez-De-Cossio2020}, characterized by order parameters $|m_w| > 0$ and $|q_w| > 0$. This analysis is confirmed after solving Equations \ref{saddle} and \ref{eq:Gaussavg}. Results are shown in Figure \ref{last}(a).
	\item For smaller values of $n$, the couplings average goes to zero, with a higher variance. It means that the couplings have a higher probability of taking different signs for different pairs of cells. This results in a disordered phase in Figure \ref{last}, with the emergence of a cross-feeding regions within the tissue, analogous to the apparition of the spin-glass phase in the disordered Ising systems \citep{Penney1993, Coolen1993, Fernandez-De-Cossio2019, Fernandez-De-Cossio2020}. In this scenario $|m_w| = 0$ and $|q_w| > 0$.
\end{itemize}

As $\mu$ increases, the dynamics evolve at a higher rate toward the fields $J_s$ (which are taken as zero), meaning that the $J_{ij}$'s follow a Gaussian distribution with mean $J_s$ and variance $\Delta \to 0$. Consequently, the system transitions to a single purely state with $|m_w| = 0$ and $|q_w| = 0$, resulting in a state where no cells secrete or absorb lactate, and the entire tissue relies on respiration.

It is important to highlight the biological significance of the existence of both ordered and disordered cooperation phases. When the average flow is zero but the second moment is non-zero, it indicates the presence of opposing flows within the culture. This is a sign of the emergence of cross-feeding cycles in the system.  If the phase is fully ordered, we are observing a case of cross-feeding that extends uniformly throughout the entire culture, which arises due to the shift toward the lactate shuttle mechanism in the cultured cells. On the other hand, the disordered or glassy phase exhibits localized regions within the culture where these cross-feeding cycles appear. Unlike the ordered case, this behavior does not propagate across the entire culture. These phenomena shows similarities to the persistence of cells in tissues (such as tumor environments) due to the reverse Warburg effect mechanism between stromal and tumoral cells.

In biological terms, these results lead to the following conclusions:
\begin{itemize}
	\item If cells are not under dominant evolutionary pressure ($\mu$ is small), the primary metabolic pathway in the tissue (respiration, fermentation, or cooperation through cross-feeding) depends on whether their couplings evolve in response to correlations between neighboring cell fluxes or in response to noise in the dynamics.
	\item If the dependence is on correlations, there is an average flow of outgoing lactate per cell (fermentation).
	\item If the dependence is primarily on noise, the average secreted lactate is zero across the tissue, but fluctuations appear in various regions where lactate is secreted and absorbed. These fluctuations correspond to the formation of cooperative cross-feeding regions.
\end{itemize}

This interpretation is validated in Figure \ref{last}(b). Introducing a $J_w < 0$ reinforces the tendency towards the disordered phase, which is consistent with the interpretation in Figure \ref{fases}.

\section{Discussion and Summary}
\label{sec:disc}

 In this work, we have developed a theoretical framework to study the emergence of metabolic interactions in microbial communities. By combining Monte Carlo simulations with analytical methods from the statistical physics of disordered systems, we explored how competitive and/or cooperative interactions arise in a population of cells. Our model highlights the relationship between glucose competition and waste exchange, revealing three distinct phases: competitive, cooperative, and disordered. These phases reflect the diverse strategies used by cells to optimize resource utilization and survival, reproducing known phenomena such as the Warburg and reverse Warburg effects in tumor tissues. The simulations performed in the two-dimensional model can be understood as an appropriate proxy for modeling biofilms or structures in simple tissues. In contrast, the analytical computations carried out in the mean-field model are more representative of a varying population of cells, such as bacteria in suspension.

From a technical point of view, we emphasize that the assumption that all cells are characterized by the same metabolic constraints is a condition imposed in this context because we were interested in modeling the behavior of isogenetic cells. However, it is not a fundamental limitation of our approach. The description of different species in a culture, each with distinct metabolic constraints, could readily be incorporated. Although this would slightly complicate the simulations and significantly increase the complexity of the analytical computations, it remains feasible within the existing formalism. Nevertheless, we note that we did not aim to fit the stoichiometric model to experimental parameters, and therefore the current model should be understood as exploratory rather than predictive. We are confident that exploring alternative distributions for interaction strengths or introducing spatial heterogeneity could offer further insights into the robustness and adaptability of microbial communities.

It is important to note that, in order to keep the analytical treatment tractable, our model does not directly account for cellular division. However, changes in coupling parameters can be interpreted as the result of the emergence of new cells with greater fitness or better adaptation to the environment. Alternatively, the impact of transcriptional changes occurring during cell division is captured via the stochastic component of the dynamical equation, modulated by the parameter $\tilde\beta$.

In a more general model, one can imagine that the spatial redistribution due to population growth could be represented through a deterministic evolutionary pressure specific to the tissue being modeled. The effects of this pressure can be absorbed into the choice of the intrinsic field $J_s$ for each tissue type. Similarly, the influence of spatially correlated cell flows can be modeled through their effects on the evolution of interaction couplings.

Imposing a Boltzmann distribution to characterize the probability space of metabolic states is often justified using the Principle of Maximum Entropy, as seen in previous works \citep{DeMartino2016,Pérez-Fernández2021,Batista-Tomás2021}. The use of an exponential distribution further facilitates parallels with methods from the statistical mechanics of disordered systems. It is important to emphasize that other feasible distributions could be employed in similar contexts, depending on the biological assumptions or modeling goals.

The main results of this work can be grouped into two fundamental parts: (i) the generalization of previously studied models that describe the equilibrium state of interacting cellular systems, and (ii) the inclusion of a dynamical process governing the interactions between cells, along with its numerical and analytical solutions. In the first part, we obtained a detailed phase diagram that maps the presence of different metabolic phases as a function of fixed interaction parameters. In particular, within the cooperative phase, we observed the emergence of opposing sub-networks where cross-feeding processes occur. Moreover, upon incorporating the evolution of interactions, we were able to study adaptive phenomena in cells located in glucose-deprived regions. We found that, in principle, these cells could survive in such hostile environments thanks to nutrients received from external neighbors. This includes the dynamic formation of cross-feeding networks, where external cells consume glucose and internal cells utilize metabolic by-products. Given that diverse cross-feeding patterns (via the reverse Warburg effect) have been observed in microbial cultures, our approach has the potential to gain biological relevance. The qualitatively similar results obtained from studying both two-dimensional and mean-field models highlight the universal applicability of the phase diagram. They also underscore that the evolutionary dynamics driving metabolic interactions are the main determinants of the phase structure, limiting the relevance of spatial system structure in explaining the observed phenomena.

Our theoretical framework also challenges the notion that cells "want to" optimize a utility function. The idea that individual cells maximize biomass or ATP production serves as a mathematical abstraction to describe complex biological processes. Similarly, our proposal—that a culture may minimize a cost function that includes intercellular interactions—should be interpreted in the same light. A simple proxy for such an interaction is the use of environmental energy sources, which may, in turn, originate from the metabolism of neighboring cells. This reasoning aligns with approaches such as those in \citep{DeMartino2017,Narayanankutty2024}, where metabolic network modeling is combined with diffusion constraints, single-cell experimental flux data, and tools from statistical physics to show that heterogeneity in a culture can be explained by metabolite exchanges among cells. Our approach opens a path to understanding how this heterogeneity arises as a result of evolutionary dynamics.

Future research could integrate empirical parameters and validate the model's predictions against real-world experimental data. Given the complexity of measuring the intrinsic parameters of the model, experimental validation can be challenging. However, the model can be adjusted to address this issue. For example, the values of the fields $h_s$ could be set to realistic levels based on existing studies of cellular metabolism. Similarly, the interaction fields $J_s$, as well as parameters such as $\beta$ and $\tilde\beta$, could be estimated using fitting procedures applied to experimental data from cellular cultures, as in \cite{Narayanankutty2024}. Furthermore, exploring alternative distributions for interaction strengths or introducing spatial heterogeneity could offer additional insights into the robustness and adaptability of microbial communities.

\section*{Acknowledgments}
This work was partially supported by Horizon 2020 Maria Sklodowska-Curie Action-Research and Innovation Staff Exchange (MSCA-RISE) 2016 (Grant Agreement No. 734439) (INFERNET: New algorithms for inference and optimization from large-scale biological data), and Ministerio de Ciencia, Tecnología y Medio Ambiente, PNCB-PN223LH010-015.

\appendix
\section{Derivation of the Analytical Solutions} 
\label{A1}

We start with the simple model described in \citep{Fernandez-De-Cossio2020} , with the corresponding Hamiltonian:
\begin{equation}\label {hamilt}
	H(\{s\})=-\sum_s\sum_i h_{r}s_{i}-\sum_s\sum_{i<j}J_{sij}s_{i}s_j
\end{equation}
where $s\in\{g,r,w\}$. The partition function of the spin system is:
\begin{equation}
	Z(\{J_{sij}\})=\underset{\{s\}}{\text{Tr}}\;e^{-\beta H(\{s \})}
\end{equation}
The flux configuration will follow an equilibrium distribution:
\begin{equation}
	P(\{s \})=\frac{e^{-\beta H(\{s \})}}{Z_\beta(\{J_{sij}\})}
\end{equation}
Now we can consider that the $J_{sij}$ follows a dynamic law with the form \citep{Coolen1993}:
\begin{equation}
	\tau\frac{d}{dt}J_{sij} =\frac{1}{N}\left\langle s_{i}s_j \right\rangle_{\{J_{sij}\}} -\mu (J_{sij}-J_s/N)+\frac{1}{\sqrt N}\eta_{sij}(t)
\end{equation}	
being $\eta_{sij}(t)$ a Gaussian noise with 0 mean and $\left\langle \eta_{sij}(t)\eta_{sij}(t')\right\rangle=\frac{2\mathcal{T}}{\tilde\beta}\Delta_{{rij},{rkl}} \Delta(t-t')$.

So, under this dynamic approach, we will consider that for each time $t$, the spins system equilibrates according to (20-22). Also, for long times, both system (spins and couplings) equilibrates, and the difference in time scales loses relevance. \\
Then, the correlations in the dynamic equation can be expressed in terms of the equilibrium distribution as:

\begin{equation}
	N\tau\frac{d}{dt}J_{sij} =-\frac{\partial \mathcal{H}(\{J_{sij}\})}{\partial J_{sij}}+ \sqrt N \;\eta_{ij}(t)\label{eq:dyn}
\end{equation}
with:
\begin{equation}
	\mathcal{H}(\{J_{sij}\})=-\frac{1}{\beta}\ln Z(\{J_{sij}\})+\frac{1}{2}\mu N\sum_{s,i<j}(J_{sij}-J_s/N)^2 
\end{equation}

There are two equilibrium processes, for $J_{sij}$ and $s$. The first one, at a slow 
pace, with associated Hamiltonian $\mathcal{H}(\{J_{sij}\})$, inverse temperature $\tilde\beta$ and partition function $Z_{\tilde\beta}$, and the other with fast pace (equilibrates for each step of the first process) with hamiltonian $H(\{J_{sij}\})$, inverse temperature $\beta$ and partition function $Z_\beta$.  The quadratic term in $\mathcal{H}$ is related to a quadratic potential centered at $J_s/N$. \\
The process \ref{eq:dyn} can be identified as a Langevin equation, with partition function:
\begin{align}
	Z_{\tilde\beta}&=\underset{\{J_{sij}\}}{\text{Tr}}\;e^{-\tilde\beta \mathcal{H}(\{s \})}=\underset{\{J_{sij}\}}{\text{Tr}}\;e^{\tilde\beta\left(\frac{1}{\beta}\ln Z(\{J_{sij}\})-\frac{1}{2} \mu N\sum_{s,i<j}\left(J_{sij}-J_s/N\right)^2 \right)}\\
	Z_{\tilde\beta}&=\underset{\{J_{sij}\}}{\text{Tr}} Z^n(\{J_{sij}\})\;e^{-\frac{1}{2}\tilde\beta \mu N\sum_{s,i<j}\left(J_{sij}-J_s/N\right)^2 }
\end{align} 
where $n=\frac{\tilde\beta}{\beta}$ and $Z^n(\{J_{sij}\})$ can be interpreted as the partition function over $n$ replicas of the same disorder configuration ($n$ is a positive integer for the moment). So:  
\begin{align}
	Z_{\tilde\beta}&=\underset{\{s\}}{\text{Tr}}\left[\underset{\{J_{sij}\}}{\text{Tr}}\left[e^{\,-\beta\sum_{a}H(v^a)}\;e^{-\frac{1}{2}\tilde\beta \mu N\sum_{s,i<j}\left(J_{sij}-J_s/N\right)^2 }\right]\right]
\end{align}
where  $a=1,2,...,n$. Calling $\Delta^2=1/\tilde\beta \mu  $ \\
\begin{align}
	Z_{\tilde\beta}=\underset{\{s_{i}^a\}}{\text{Tr}}\int \prod_{i<j}\prod_{s}dJ_{sij}\text{exp}\left[-\frac{N}{2\Delta^2}\left(J_{sij}-J_s/N\right)^2\right]\,\text{exp}\left(\beta \sum\limits_a\sum_{i} h_i s_{i}^a+\beta \sum\limits_a J_{sij} s_{i}^a s_j^a\right)
\end{align}

Integrating and neglecting subexponential term as $N\to \infty$, we obtain:
\begin{gather}
	Z_{\tilde\beta}=\underset{\{s_{i}^a\}}{\text{Tr}}\prod_{s}\text{exp} \left[\beta \sum\limits_{i,a} h_i s_{i}^a+\frac{\beta J_s }{N}\sum_{i<j}\sum_{a} s_{i}^a s_j^a+\sum_{i<j}\frac{\beta^2 \Delta^2  }{2N}\left(\sum_{a} s_{i}^a s_j^a\right)^2\right]\\
	=\underset{\{s_{i}^a\}}{\text{Tr}}\prod_{s}\text{exp} \left[\beta \sum\limits_{i,a} h_i s_{i}^a+ \sum_{a}\frac{\beta J_s}{2N}\left(\sum_{i} s_{i}^a\right)^2 +\sum_{a\le b}\frac{\beta^2\Delta^{2}}{2N}  \left(\sum_{i} s_{i}^a s_{i}^b\right)^2\left(1-\frac{\Delta_{a b}}{2}\right)\right]\nonumber
\end{gather}
Now, we introduce the variables $m^a_r=\sum_{i} s_{i}^a\;,\;q^{ab}_r=\sum_{i} s_{i}^a s_{i}^b$ and apply the Hubbard-Stratonovich \\ $\left(e^{y x^2/N}=\sqrt{\frac{N}{2\pi}}\int e^{-(N/2) u^2+\sqrt{2y}x u }du\right)$ to eliminate the squared terms in the exponent:
\begin{gather}
	Z_{\tilde\beta}=\underset{\{s_{i}^a\}}{\text{Tr}}\prod_{s}\int\left(\prod_{a}\sqrt\frac{N}{2\pi}dm^a_r\right)\left(\prod_{a\le b}\sqrt\frac{N}{2\pi}dq^{a  b}_r\right)\text{exp}\left(-\sum_a\frac{N}{2}(m^a_r)^2-\sum_{a\le  b}\frac{N}{2}(q^{a b}_r)^2\right)\times\\
	\times \prod_{i}\text{exp}\left(\beta \sum\limits_{a} h_i s_{i}^a+\sum\limits_{a}\sqrt{\beta J_s}m^a_rs_{i}^a+\sum\limits_{a\le b}\frac{\beta \Delta q^{a b}_r}{\sqrt{1+\Delta_{a b}}} s_{i}^a s_{i}^b\right)\nonumber
\end{gather}

Note that the trace acts only over a single cell site. Then, and neglecting subexponential terms again, the partition function becomes:
\begin{gather}
	Z_{\tilde\beta}=\int\prod_{s}\left(\prod_{a}dm^a_r\prod_{a\le b}dq^{a b}_r\right)\text{exp}\left(-\sum_a\frac{N}{2}(m^a_r)^2-\sum_{a\le  b}\frac{N}{2}(q^{a b}_r)^2\right)\times\nonumber\\
	\times \text{exp}\left[N\ln \underset{\{s^a\}}{\text{Tr}}\prod_{s}\text{exp}\left( \sum\limits_{a}\left( \beta h_s+ \sqrt{\beta J_s}m^a_r\right)s^a+\sum\limits_{a\le b}\frac{\beta \Delta q^{a b}_r}{\sqrt{1+\Delta_{a b}}} s^a s^b\right)\right]\\
	Z_{\tilde\beta}=\int\prod_{s}\left(\prod_{a}dm^a_r\prod_{a\le b}dq^{a b}_r\right)\text{exp}\left(Nnf(m^a_r,q^{a b}_r)\right)\approx\text{exp}\left(Nnf(m^{*a}_r,q^{*a b}_r)\right)
\end{gather}
where:
\begin{align}\label{f_1}
	f(m^{a}_r,q^{a b}_r)&=-\sum_{a,r}\frac{(m^{a}_r)^2}{2n}-\sum_{a\le b,r}\frac{(q^{a b}_r)^2}{2n}+ \nonumber\\
	&+\frac{1}{n}\ln \underset{\{s^a\}}{\text{Tr}}\prod_{s}\text{exp}\left( \sum\limits_{a}\left(\beta h_s+ \sqrt{\beta J_s}m^{a}_r\right)s^a+\sum\limits_{a\le b}\frac{\beta \Delta q^{a b}_r}{\sqrt{1+\Delta_{a b}}} s^a s^b\right)
\end{align}
is the average free energy function for one replica and cells for $N$ large, and $m^{a*}_r$ and $q^{*a b}_r$ are such that they extremize  $f(m^a_r,q^{a b}_r)$. 	

Now, we take the replica symmetric anzats, where $m^{a}_s=m_s$, $q^{ab}_s=q_s$ ($a<b$) and $q^{aa}_s=\zeta_s$. Using the transformation:
\begin{align}
	m_s\to \frac{m_s}{\sqrt{\beta J_s}}\;\;\;\;\;,\;\;\;\;\;q_s\to\frac{q_s}{\beta \Delta}\;\;\;\;\;,\;\;\;\;\;
	\zeta_s\to\frac{\zeta_s\sqrt 2}{\beta \Delta}
\end{align} 

we obtain:
\begin{align}
	f(m^a_r,q^{a b}_r)&=-\sum_s\left[\frac{\beta J_s m_s^2}{2}+\frac{\beta^2\Delta^{2}}{4}\Big(\zeta_s^2+(n-1)q_s^2\Big)\right]+\nonumber\\
	&+ \frac{1}{n}\ln \text{Tr}\prod_{s}\text{exp}\left( \sum\limits_{a}\left( \beta h_s+ \beta J_s m_s\right)s^a+\beta^2\Delta^{2}q_s\sum\limits_{a< b} s^a s^b+\frac{\beta^2\Delta^{2}\zeta_s}{2}\sum\limits_{a} (s^a)^2 \right)\nonumber\label{f_2}
\end{align}
Using $\sum_{a<b}s^as^b=\frac{1}{2}(\sum_as^a)^2-\sum_{a}(s^a)^2$ and the H-S transform, introducing the variable $t_s$ we eliminate the squared sums in the exponent. Now, the trace runs over a single replica and:
\begin{align}
	f(m^a_r,q^{a b}_r)&=-\sum_s\left[\frac{J_s m_s^2}{2}+\frac{\Delta^{2}}{4}\Big(\zeta_s^2+(n-1)q_s^2\Big)\right]+\nonumber\\
	&+ \frac{1}{n}\ln \int \left\{\text{Tr}\prod_{s} \text{exp}\left[\left(\beta h_s+ \beta J_s m_s+\beta \Delta\sqrt{q_s}t_s\right)s^a+\frac{\beta^2\Delta^2}{2}(\zeta_s-q_s)(s^a)^2\right] \right\}^n \vec {Dt_s}\nonumber\\
	\nonumber\\
	f(m^a_r,q^{a b}_r)&=-\sum_s\left[\frac{ J_s m_s^2}{2}+\frac{\Delta^{2}}{4}\Big(\zeta_s^2+(n-1)q_s^2\Big)\right]+ \frac{1}{n}\ln \int \left\{\text{Tr }\left[\text{exp } (-\mathbb{H}_{RS})\right] \right\}^n\vec{Dt_s}\nonumber
\end{align}
where we have called $\mathbb{H}_{RS}=-\sum_s\left(\beta h_s+ \beta J_s m_s+\beta \Delta \sqrt{q_s}t_s\right)s-\sum_s\frac{\beta^2\Delta^{2}}{2}(\zeta_s-q_s) (s)^2$. Using the saddle point conditions gives:
\begin{gather}
	m_s^*= \frac{\displaystyle\int\left [\text{Tr }e^{-\mathbb{ H}_{RS}}\right]^n\left(\frac{\text{Tr }e^{-\mathbb{ H}_{RS}}s}{\text{Tr }e^{-\mathbb{ H}_{RS}}}\right) D\vec t_s}{\displaystyle\int\left [\text{Tr }e^{-\mathbb{ H}_{RS}} \right]^n D\vec t_s}\\\nonumber\\
	q_s^*=\frac{\displaystyle\int\left [\text{Tr }e^{-\mathbb{ H}_{RS}}\right]^n\left(\frac{\text{Tr }e^{-\mathbb{ H}_{RS}}s}{\text{Tr }e^{-\mathbb{ H}_{RS}}}\right)^2 D\vec t_s}{\displaystyle\int\left [\text{Tr }e^{-\mathbb{ H}_{RS}} \right]^n D\vec t_s}\\\nonumber\\
	\zeta_s^*=\frac{\displaystyle\int\left [\text{Tr }e^{-\mathbb{ H}_{RS}}\right]^n\left(\frac{\text{Tr }e^{-\mathbb{ H}_{RS}}s^2}{\text{Tr }e^{-\mathbb{ H}_{RS}}}\right) D\vec t_s}{\displaystyle\int\left [\text{Tr }e^{-\mathbb{ H}_{RS}} \right]^n D\vec t_s}
\end{gather}
where $D\vec t_s = \prod_s\frac{1}{\sqrt{2\pi}}e^{\frac{-t_s^2}{2}}dt_s$ 

The equilibrium solution (without the coupling dynamics) for each instant can be obtained after taking the limit $n\to 0$, while maintaining a fixed $\beta$. With that, it is:
\begin{align}
	m_s^*= \displaystyle\int\frac{\text{Tr }e^{-\mathbb{ H}_{RS}}s}{\text{Tr }e^{-\mathbb{ H}_{RS}}} D\vec t_s\;\;\;\;,\;\;\;\;\;
	q_s^*=\displaystyle\int\left(\frac{\text{Tr }e^{-\mathbb{ H}_{RS}}s}{\text{Tr }e^{-\mathbb{ H}_{RS}}}\right)^2 D\vec t_s\;\;\;,\;\;\;
	\zeta_s^*=\displaystyle\int\frac{\text{Tr }e^{-\mathbb{ H}_{RS}}s^2}{\text{Tr }e^{-\mathbb{ H}_{RS}}} D\vec t_s
\end{align}

\section{Table of Symbols (in order of appearance)}
\label{A2}
\begin{table}[h!]
\centering
\begin{tabular}{ll}
\hline
\textbf{Symbol} & \textbf{Description} \\
\hline
$S$ & Substrate taken from the environment \\
$E$ & Energy obtained from metabolism \\
$P$ & Pyruvate formed after glycolysis \\
$W$ & Waste by product secreted after fermentation \\
$s=\{g,r,w\}$ & Reaction fluxes from glucose absorption, respiration, and waste secretion/absorption \\
$\textbf{lb}_g,\textbf{ub}_r$ & Lower ad upper bounds of fluxes $g$ and $r$ respectively  \\
$E_i$ & Utility function of cell $i$, directly related to biomass production \\
$h_s$ & Contribution from reaction $s$ to the utility function \\
$V$ & Contribution to the utility function from interactions between neighbors \\
$H$ & It is defined in a similar way as the Hamiltonian $H=\sum_iE_i+V$ \\
$J_sij$ & Coupling strength between neighbors $i,j$ through the reaction $s$ \\
$\beta$ & "Inverse temperature" related to the spin variables (fluxes)  \\
$J_s$ & Intrinsic value for all cells' couplings (through reaction $s$) to evolve in the absence of correlations\\
$\mu$ & Rate at which couplings evolve towards $J_s$ \\
$\eta_{sij}$ & Noise associated to the evolution of coupling $J_{sij}$ \\
$\tau$ & Timescale separating the fluxes and coupling dynamics. It is also the time covariance of the coupling noise. \\
$\tilde \beta$ & Inverse temperature of the couplings' dynamics. \\
$\Delta^2$ &Variance of the distribution at which cuplings evolve in the absence of correlations   \\
$n$ & Ratio between $\tilde\beta$ and $\beta$. Separates the timescales between dynamics of the fluxes and couplings \\
$m_s$ & Average of the flux $s$ across all the cells across all replicas.  \\
$q_s$ & Average overlap of the flux $s$ between neighboring cells across all replicas \\
$\zeta_s$ & Overlap $s$ between neighboring cells of the same replica, averaged across all replicas\\

\hline
\end{tabular}
\caption{List of symbols used.}
\end{table}

\end{document}